# AC thermal conductivity as a tool for solution mapping from diffusive to ballistic regime

Tao Li, Bo Jiang, and Zhen Chen*

School of Mechanical Engineering, Southeast University, Nanjing 210096, China

* To whom correspondence should be addressed: zhenchen@seu.edu.cn

**Abstract**

Although the Boltzmann transport equation (BTE) has been exploited to investigate non-diffusive phonon transport for decades, due to the challenges of solving this integro-differential equation, most standard techniques for thermal measurements still rely on solutions to the diffusion equation, causing inconsistency between measured non-diffusive effects and the diffusion equation based techniques. With the AC thermal conductivity, an analogous concept of the AC electrical conductivity in solid state physics, we transform BTE under the relaxation time approximation into the form of the diffusion equation. This transformation maps any analytical solution of the diffusion equation under periodic heating to that of the BTE, with the nonlocal effect captured by the jump boundary condition. After investigating the validity of this framework, we apply it to generalize the $3\omega$ method from diffusive to quasi-ballistic, and propose an experimental scheme to address the inconsistency problem above.

## I. Introduction

The frequency-dependent electrical conductivity, also known as AC conductivity ($\sigma_{AC}$), is an important concept in solid state physics.[1] This concept describes that the driving force, the electrical field in this case, oscillates so fast, that the electrons cannot respond instantaneously. As a result, a phase lag appears between the electrical current and the electrical field. This phase lag, and correspondingly $\sigma_{AC}$, is the origin of many important phenomena, such as the well-known plasma frequency.[1]

An analogous concept in the thermal domain is the AC thermal conductivity, $k_{AC}$ (Appendix A).[2-5] Likewise, this $k_{AC}$ manifests the thermal inertial effects, which causes a similar phase lag between the heat flux and the temperature field. Although this $k_{AC}$ has been proposed, existing literature was limited to the computation of $k_{AC}$ itself.[2-5]

On the other hand, non-diffusive thermal transport has been investigated both theoretically and experimentally for decades.[6-9] However, the widely used experimental techniques, e.g. the 3ω method[10-11] and the time/frequency domain thermal reflectance method,[12-14] rely on solutions to the diffusion equation, leading to inconsistency between measured non-Fourier effects and the Fourier-law based techniques.[7-8]

The phonon Boltzmann transport equation (BTE) can deal with non-diffusive transport.[15-18] However, due to the seven-variable dependence of this integro-differential equation, solving BTE analytically for a general space-time-dependent problem, which is preferred in experimental analyses, remains a challenging task, even under the relaxation time approximation (RTA). Although there are precedent attempts to circumvent the inconsistency above analytically,[19-24] these studies either neglect the nonlocal effect,[19] or are restricted to (semi-) infinite domains,[20-21] one-dimensional problems,[22-23] or gray models.[22, 24] Formalisms relying on equations other than Boltzmann transport, e.g. the truncated Levy[25] or the hydrodynamic equation,[26] have also been proposed to directly fit for thermal properties from experimental raw data.

Here for periodic heating problems of phonon transport, we apply the phonon-photon analogy to develop a general framework to obtain analytical solutions to BTE under RTA. This framework casts BTE into the diffusion equation, in which the conventional real thermal conductivity, $k_{bulk}$, is replaced by its complex counterpart, $k_{AC}$. Analogous to the treatment in classic radiative transport, we apply the Deissler jump boundary conditions to take into account the nonlocal effect.[27-28] This framework maps any existing general solution (with undetermined coefficients) of the diffusion equation to that of the BTE.

We first investigate the validity of this framework by comparison to both analytical solutions in the limiting cases and numerical solutions of a gray model. Then we generalize the heat transfer model of the classic 3ω method to be able to capture quasi-ballistic phonon transport. We show that, in some cases, the commonly used $k_{eff}$-approach, which fits experimental raw data to a diffusion solution with an effective thermal conductivity, is not adequate to describe the full temperature distribution. Correspondingly, we suggest new experimental scheme, and demonstrate using synthetic experiments.

## II. Framework

We exploit the close analogy between phonon transport and photon radiation to develop both the governing equation and the boundary conditions. For simplicity, we assume the phonon dispersion and the phonon scattering are both isotropic, reserving the anisotropic cases[29-30] for future developments.

### A. Governing equation

With the phonon-photon analogy, the Boltzmann transport equation (BTE) under the single mode relaxation time approximation (RTA) is transformed to the equation of phonon radiative transfer,[15]

$$\frac{\partial I_\omega}{\partial t} + v_\omega \hat{\mathbf{s}} \cdot \nabla I_\omega = -\frac{I_\omega - I_\omega^0}{\tau_\omega}, \tag{1}$$

where $\omega$ is the angular phonon frequency, $v_\omega$ is the magnitude of the phonon group velocity, $\hat{\mathbf{s}}$ is the unit vector in the direction of the velocity, and $\tau_\omega$ is the relaxation time. The phonon intensity, $I_\omega$, is linked to the phonon distribution function, $f_\omega$, through[15]

$$I_\omega = v_\omega \hbar \omega D(\omega) f_\omega / 4\pi, \tag{2}$$

where $\hbar$ is the reduced Plank's constant, and $D(\omega)$ is the phonon density of states (DOS). Therefore, $I_\omega^0$ corresponds to the equilibrium distribution function, $f_\omega^0$.

Following the Milne-Eddington approach,[27] i.e., multiplying Eq. 1 by the unit vector, $\hat{\mathbf{s}}$, and then integrating it over the entire $4\pi$ solid angle, one obtains

$$\frac{\partial q_{net,\omega}}{\partial t} + \frac{v_\omega}{3} \cdot \nabla G_\omega = -\frac{q_{net,\omega}}{\tau_\omega}, \tag{3}$$

where $\mathbf{q}_{net,\omega} = \int_{4\pi} \hat{\mathbf{s}} \cdot I_\omega d\Omega$ and $G_\omega = \int_{4\pi} I_\omega d\Omega$ are the mode-wise net radiative phonon flux and incident phonon radiation. With Eq. 2, one arrives at $G_\omega = \|\mathbf{v}_{g,\omega}\| \cdot e_\omega$, where $e_\omega$ is the mode-wise internal energy. Note here to obtain Eq. 3, we invoke the lowest-order approximation of the spherical harmonics method (also known as the P$_1$ approximation),[27]

$$I_\omega(\mathbf{r}) \approx \frac{1}{4\pi}[G_\omega(\mathbf{r}) + \hat{\mathbf{s}} \cdot 3\mathbf{q}_{net,\omega}(\mathbf{r})], \tag{4}$$

which assumes phonon intensities has two terms: the first term is independent of direction and related to the spectral incident radiation $G_\omega$, and the second is related to the *x, y, z*-component of the spectral net heat flux $\boldsymbol{q}_{net,\omega}$ vector respectively.

We combine Eq. 3 and the mode-wise energy conservation,

$$\frac{\partial G_\omega}{\partial t} + v_\omega \cdot \nabla \cdot \boldsymbol{q}_{net,\omega} = 0, \tag{5}$$

to eliminate $\boldsymbol{q}_{net,\omega}$ and obtain a hyperbolic (or, a telegraph) equation,[6, 31]

$$\frac{\partial^2 G_\omega}{\partial t^2} + \frac{1}{\tau_\omega}\frac{\partial G_\omega}{\partial t} = \frac{v_\omega^2}{3}\nabla^2 G_\omega. \tag{6}$$

Note here Eq. 6 shares the same form with the equation describing the wave propagation in media of finite electrical conductivity,[27] which is not a typical wave equation but includes an extra first-order derivative term with respect to time. In that scenario, one introduces a complex refractive index, and thus reduces the equation to the typical wave equation. In the following, motivated by this treatment, we are able to transform Eq. 6 to the familiar diffusion-type equation (Appendix B).

Our framework so far is general for any transient problem. In the following, we now restrict our analysis to periodic heating problems. According to the Linear Response Theorem, the thermal response of a linear system to a periodic heating excitation with an angular frequency of $\omega_H$ reads as $G_\omega(\mathbf{r},t) = \bar{G}_\omega(\mathbf{r})e^{i\omega_H t}$ and $\boldsymbol{q}_{net,\omega}(\mathbf{r},t) = \bar{\boldsymbol{q}}_{net,\omega}(\mathbf{r})e^{i\omega_H t}$. Substituting these two expressions to Eqs. 3 and 6, one obtains

$$\frac{1}{\alpha_{AC,\omega}}\frac{\partial G_\omega}{\partial t} = \nabla^2 G_\omega, \tag{7}$$

$$\boldsymbol{q}_{net,\omega} = -\frac{k_{AC,\omega}}{C_\omega v_\omega} \cdot \nabla G_\omega, \tag{8}$$

where $\alpha_{AC,\omega} = k_{AC,\omega}/C_\omega$ is the spectral diffusivity, in which we define a spectral AC thermal conductivity,

$$k_{AC,\omega} = \frac{C_\omega v_\omega^2 \tau_\omega/3}{1+i\omega_H \tau_\omega}. \tag{9}$$

Note the factor, $1 + i\omega_H\tau_\omega$, in the denominator is a result of the $e^{i\omega_H t}$ notation, which should be replaced to $1 - i\omega_H\tau_\omega$ if we use an $e^{-i\omega_H t}$ notation.

Equations 7-9 are the most important results of the article. With the AC thermal conductivity (Eq. 9) and the P$_1$ approximation (Eq. 4), the phonon BTE (Eq. 1) under periodic heating is cast into the form of the familiar diffusion equation (Eqs. 7-9). Consequently, for any existing analytical solution to the diffusion equation under periodic heating, as summarized in classic textbooks, e.g. by Carslaw and Jaeger,[32] one simply replaces $k_{bulk}$ with $k_{AC,\omega}$ to obtain the corresponding general solution of the phonon BTE (with undetermined coefficients), because the same equations have the same solutions.

One caveat, though, is that the P1 approximation (Eq. 4) transforms the nonlocal[33] phonon BTE (Eq. 1) into a local diffusion-type equation (Eq. 8). This locality, i.e., the flux at a point, **r**, is entirely determined by the gradient of the incident phonon radiation at the same point, **r** (Eq. 7), limits the capability of the original equation to describe nonlocal transport effects. To address this shortcoming, we follow the treatment in radiative transport[27] to capture the nonlocal transport effects with the modified boundary conditions. As a result, although Eq. 8 and the classical diffusion equation share the same general form of solutions, the undetermined coefficients are different due to the modified boundary conditions, as described in next section and manifested in Table I for three representative examples.

**B. Boundary conditions**

To take into account the nonlocal effect, we continue with the analogy between phonon transport and traditional photon radiative heat transfer, and invoke the Deissler jump boundary conditions.[27-28] In the following, we outline the key steps to arrive at these boundary conditions.

Analyzing the energy balance at the boundary, one obtains[27]

$$E_{b,\omega,bdy} - J_\omega = \frac{1-\epsilon_\omega}{\epsilon_\omega} \boldsymbol{q}_{net,\omega,bdy} \cdot \hat{\mathbf{n}} \tag{10}$$

where $\epsilon_\omega$ is the emissivity, and $\hat{\mathbf{n}}$ is the unit vector of the surface normal of the boundary. In the context of thermal measurement of phonon transport, e.g. the $3\omega$ method in Section V, $\epsilon_\omega$ can be interpreted as the transmission coefficient from the heater to the sample.[6, 24] $E_{b,\omega}$ and $\boldsymbol{q}_{net,\omega,bdy}$ are the spectral emissive power and net flux at the boundary, respectively. The spectral radiosity is defined as the spectral energy flux outgoing from an infinitesimal surface, $J_\omega = \int_{\hat{\mathbf{s}}\cdot\hat{\mathbf{n}}>0} I_\omega \hat{\mathbf{s}} \cdot \hat{\mathbf{n}} d\Omega$. Likewise, the spectral incident energy flux, known as the spectral irradiation, is defined over the incident hemisphere, $H_\omega = \int_{\hat{\mathbf{s}}\cdot\hat{\mathbf{n}}<0} I_\omega \hat{\mathbf{s}} \cdot \hat{\mathbf{n}} d\Omega$. With the P1 approximation (Eq. 4), one arrives at

$$J_\omega = \frac{1}{4}\left(G_{\omega,bdy} + 2\boldsymbol{q}_{net,\omega,bdy} \cdot \hat{\mathbf{n}}\right), \tag{11a}$$

$$H_\omega = \frac{1}{4}\left(G_{\omega,bdy} - 2\boldsymbol{q}_{net,\omega,bdy} \cdot \hat{\mathbf{n}}\right). \tag{11b}$$

We note that Eqs. 11a-b can also be derived under the Schuster-Schwarzschild approximation, also known as the two-flux model.[27, 34]

Substituting Eq. 11a into Eq. 10 to eliminate $J_\omega$, one obtains the jump boundary condition,[27-28]

$$G_{\omega,bdy} = 4E_{b,\omega,bdy} - \frac{2(2-\epsilon_\omega)}{\epsilon_\omega}\boldsymbol{q}_{net,\omega,bdy} \cdot \hat{\mathbf{n}}, \tag{12}$$

where the second term on the right-hand side introduces the jump boundary condition, which is a signature of the ballistic transport.[22]

In the following, we discuss the temperature and flux boundary conditions, respectively.

1. **Temperature boundary condition**

We assume a periodic temperature at the boundary, $T_{bdy}(\mathbf{r},t) = \bar{T}_{bdy}(\mathbf{r})e^{i\omega_H t}$, as the driving force. Substituting Eq. 7 into Eq. 12 to eliminate $q_{net,\omega}$ and the common factor $e^{i\omega_H t}$, one obtains

$$4\bar{E}_{b,\omega,bdy} = \bar{G}_{\omega,bdy} - \frac{2(2-\epsilon_\omega)}{\epsilon_\omega}\frac{k_{AC,\omega}}{C_\omega v_\omega}\nabla\bar{G}_{\omega,bdy}\cdot\hat{\mathbf{n}}, \tag{13}$$

at the boundary, where $E_{b,\omega,bdy}(\mathbf{r},t) = \bar{E}_{b,\omega,bdy}(\mathbf{r})e^{i\omega_H t}$, $G_{\omega,bdy}(\mathbf{r},t) = \bar{G}_{\omega,bdy}(\mathbf{r})e^{i\omega_H t}$, and $q_{net,\omega,bdy}(\mathbf{r},t) = \bar{q}_{net,\omega,bdy}(\mathbf{r})e^{i\omega_H t}$, according to the Linear Response Theorem. Equation 13 links the imposed boundary temperature, $\bar{T}_{bdy}$, to the incident radiation, $\bar{G}_\omega$, at the boundary through the linearized emissive power,[35]

$$\bar{E}_{b,\omega,bdy} = \frac{1}{4}C_\omega v_\omega(\bar{T}_{bdy} - T_{ref.}), \tag{14}$$

where $C_\omega$ and $v_\omega$ are the spectral specific heat and group velocity, and $T_{ref.}$ is a constant reference temperature.[35]

2. **Flux boundary condition**

If a periodic spectral flux, $q_{net,\omega,bdy} = \bar{q}_{net,\omega,bdy}e^{i\omega_H t}$, is imposed at the boundary, we obtain directly from Eq. 8 that

$$\bar{q}_{net,\omega,bdy}\cdot\hat{\mathbf{n}} = -\frac{k_{AC,\omega}}{C_\omega v_\omega}\cdot\nabla\bar{G}_{\omega,bdy}, \tag{15}$$

which bridges the flux boundary condition, $\bar{q}_{net,\omega,bdy}$, and the incident phonon radiation, $\bar{G}_{\omega,bdy}$, to bound the governing equation (Eq. 7).

We note that the temperature and flux boundary conditions are equivalent.[19] For example, once a temperature boundary condition, $\bar{T}_{bdy}$, is specified, one immediately obtains $\bar{E}_{b,\omega,bdy}$ (Eq. 14) and thus $\bar{G}_{\omega,bdy}$ (Eq. 13), which finally links to $\bar{q}_{net,\omega,bdy}$ (Eq. 15).

In practical experiments, however, instead of the spectral flux, $\bar{q}_{net,\omega,bdy}$, the total flux, $\bar{q}_{net,bdy} = \int_0^\infty \bar{q}_{net,\omega,bdy}d\omega$, is usually specified. We reserve this more complicated scenario till Session IV in the context of the ballistic $3\omega$ method.

## C. Solution strategy

We now combine the governing equation (Eq. 7) and the boundary conditions (Eqs. 13 and 15), and suggest a solution strategy. First, we search existing analytical solutions to the diffusion equation under periodic heating. If such solutions exist, as outlined in Table I for three representative examples, we simply replace $k_{bulk}$ in the solution of the temperature field, $T$, with $k_{AC,\omega}$ to obtain the general solution of the incident radiation, $\bar{G}_\omega$; if not, we solve Eq. 7 to obtain the general solution of $\bar{G}_\omega$. Next, we apply the boundary conditions (Eq. 13 or 15) to determine the coefficients in the general solution of $\bar{G}_\omega$. After $\bar{G}_\omega$ is solved, we apply Eq. 15 to obtain $\bar{q}_{net,bdy}$. At last, we use the jump boundary condition (Eq. 13) and the linearized relation (Eq. 14) to obtain the temperature, $\bar{T}_{bdy}$. Figure 1 shows an example of implementing the framework to generalize the classic $3\omega$ method from diffusive to ballistic (see Section V for details), in which the similarity of the governing equations and the jump boundary condition are both highlighted.

## III. Validity of the framework

It is well-known that the P$_1$-approximation for steady-state radiation problem breaks down for large Knudsen numbers ($Kn_{gray} = \Lambda_{gray}/L_c \gg 1$), in particular for multidimensional geometries in cylindrical or spherical coordinate.[27] We will show that the periodic-heating problems inherit this weakness, which, however, could be significantly improved in the high heating-frequency regime.

In the following, we will first show that our analytical solutions can recover existing ones at limiting cases, and then we will use the spherical geometry as an example to investigate the validity of our framework by comparison to numerical solutions.

## A. Verification by analytical solutions in limiting cases

In Table I, we apply our general framework in Section II to three typical examples in the cartesian, cylindrical, and spherical coordinate, respectively, with well-known analytical Fourier-law solutions. Since the temperature and the flux boundary conditions are equivalent, as we commented in Section II-B-2, here we assign the former for these examples, deferring the latter till we develop the ballistic $3\omega$ method in Section V. For clarity, we omit the subscript, $\omega$, from the corresponding variables in this section.

As indicated in the schematics of Table I, the boundary conditions for the Fourier-law problems are set to be $T_1$ and $T_2$, respectively, where $T_1 = \Delta T e^{i\omega_H t} + T_2$, while those for the corresponding BTE problems are $E_{b,1}$ and $E_{b,2}$, where $E_{b,1} = \Delta E_b e^{i\omega_H t} + E_{b,2}$. To ensure a compact form of these solutions, we further define dimensionless variables, $\bar{T}^* = (\bar{T} - T_c)/\Delta T$ and $\bar{G}^* = (\bar{G} - 4E_{b,c})/4\Delta E_b$.

Comparing the Fourier and the BTE solutions, we reiterate the key features of the BTE solutions in this work. First, the BTE solutions (right column) shares the same mathematical form with the Fourier solutions (left column) for corresponding problems, except that $k_{bulk}$ is replaced by $k_{AC}$ (Eq. 9), which takes into account of the heating-frequency effect and is embedded in $u_{AC} = \sqrt{i\omega_H/\alpha_{AC}}$. This is because the same equations, Eq. 7 and the classic diffusion equation, have the same solutions, as we emphasized throughout this manuscript. Second, the coefficients, $C_{geom.,n}$, of the BTE solutions are quite different from those of the Fourier-law solutions for corresponding problems, due to the jump boundary conditions (Eq. 13). Note here the 1st subscript of the coefficients, $geom. = p, c, s$, indicates the planar, cylindrical, and spherical geometries, respectively, and the 2nd subscript, $n$, distinguishes various coefficients of the same geometry. We outline the expressions of these coefficients in Table S1 of Appendix C.

We next confirm that our BTE solutions reduce to those of the semi-infinite scenarios,[19, 24, 36] when the characteristic length is much larger than the penetration depth, $L_c \gg Lp_{BTE}$. Here[19, 24]

$$Lp_{BTE} = 1/Re(u_{AC}) = Lp_{FL}/\sqrt{-\gamma + \sqrt{\gamma^2 + 1}}, \quad (16)$$

where $\gamma = \omega_H \tau$, and $Lp_{FL} = \sqrt{2\alpha_{bulk}/\omega_H}$. We use the infinite parallel plates (1st row of Table I), where $L_c = L$, as an example. In this case, we can simplify the BTE solution further to be $\Delta \bar{T}_{BTE,equil.} = \frac{\int_0^\infty \bar{G}(x)d\omega}{\int_0^\infty Cv d\omega} = \Delta T \int_0^\infty \frac{Cv}{1+2\frac{2-\epsilon}{\epsilon}\frac{k_{AC}u_{AC}}{Cv}} e^{-u_{AC}x} d\omega /$

$\int_0^\infty Cv\, d\omega$, which recovers the seemingly complicated Eq. 11 of Ref. 19. Likewise, the apparently complicated suppression function to the bulk thermal conductivity in Eq. 5 of Ref. 19 can be significantly simplified to the compact expression of Eq. 9 in this article (Appendix D).

At last, we also confirm that our BTE solutions successfully recover their corresponding Fourier solutions in the low-frequency ($\gamma \ll 1$) and/or large-size ($Kn = \Lambda/L_c \ll 1$) limit. Again, we use the infinite parallel plates (1st row of Table I) to illustrate this point. In the low-frequency limit ($\gamma \ll 1$), the intermediate variable, $\mu_{AC} = k_{AC}u_{AC}/Cv = \sqrt{i\gamma/3(1+i\gamma)}$, approaches to zero, and thus both $C_{p,1}$ and $C_{p,2}$ approach to one, as is evident from the 1st row of Table S1 in Appendix C. Therefore, the expression of the BTE solution (right column, 1st row of Table I) reduces to that of the corresponding Fourier solution (left column).

## B.  Comparison to numerical results of the gray model

Since the planar problem has been numerically verified in Ref. 19, we will focus on periodic heating problems in cylindrical and spherical coordinate, in which there are three characteristic lengths: $r_1$, $\Delta r = r_2 - r_1$, and $Lp_{BTE}$ (corresponding to $\gamma_{gray}$ according to Eq. 16). While the second option, $L_c = \Delta r$, has been widely used for steady-state problems in textbooks,[27] most popular experimental techniques, such as the $3\omega$ method[10-11] that will be discussed in next section, assume semi-infinite domains. Therefore, for simplicity we will investigate the validity of this framework as a function of two dimensionless parameters, $Kn_{gray} = \Lambda_{gray}/r_1$ and $\gamma_{gray} = \omega_H \tau_{gray}$, in which the latter is equivalent to, but more convenient than, $Kn'_{gray} = \Lambda_{gray}/Lp_{BTE}$.

Using the concentric spheres as a representative example, Fig. 2 compares our analytical results (red lines) to their corresponding numerical results (symbols), with the Fourier-law results as references (black dashed lines). In the numerical scheme, we follow Ref. 15-17 to solve the transient gray BTE, We set large enough $\Delta r$ (as compared to the penetration depth, $Lp_{BTE}$) to ensure semi-infinite domain, divide the spatial and angular spaces into small enough intervals to guarantee convergence of simulations, choose small enough time step, $\Delta t$, to assure stable simulations, and long enough total simulation time, $t_{max}$, to make sure the system to reach quasi steady-state. We plot the dimensionless incident radiation, $\bar{G}^*(r_1) = (\bar{G}(r_1) - 4E_{b,2})/4\Delta E_b$ (1st row), and the dimensionless phonon net flux, $\bar{q}^*_{net}(r_1) = \bar{q}_{net}(r_1)/Cv\Delta T$ (2nd row), as a function of $\gamma_{gray}$, for $Kn_{gray} = 0.1$ and 1 (columns), respectively. Note here, in the Fourier-law results, we define $\bar{G}^*_{FL}(r_1) = \frac{Cv\bar{T}(r_1) - CvT_2}{Cv\Delta T} = \frac{\bar{T}(r_1) - T_2}{\Delta T}$.

Figure 2 shows several distinct features. First, for small-$Kn_{gray}$ (e.g. $Kn_{gray} = 0.1$ in Fig. 2a), our analytical results (red lines) agree well with their corresponding numerical results (symbols), both of which converge to the Fourier-law results (black dashed lines) in the small-$\gamma_{gray}$ limit. Second, as $Kn_{gray}$ increases (e.g. $Kn_{gray} = 1$ in Fig. 2b), our analytical results start to deviate from the numerical results, in particular in the small $\gamma_{gray}$ regime. However, the discrepancy reduces as the increase of $\gamma_{gray}$. This is because phonons are localized to the vicinity of the inner sphere at high heating-frequencies, and thus the heat transfer pattern transforms to quasi-planar heating. The analytical solutions of this ballistic planar-heating had been verified to agree well with the numerical results.[19] Third, in the large-$\gamma_{gray}$ limit as shown in both Figs. 2a and 2b, both the analytical and the numerical results successfully recover the large-$Kn$ limit of thermal radiation: $\bar{G}^* = 0.5$ and $\bar{q}^*_{net} = 0.25$.[27] At last, the discrepancy between the analytical and the numerical solutions will increase as $Kn_{gray}$ increases further to be beyond unity, which, however, is less common in practical experiments. In addition, in this ultra-small size regime, the Boltzmann transport equation, which is based on the particle description of phonons, is likely to fail anyway.

## V. A quasi-ballistic $3\omega$ method

With the framework developed and validated above, we first generalize the classic $3\omega$ method to capture the quasi-ballistic effect. In particular, in addition to the two types of boundary conditions discussed in II-B, we consider a more realistic flux boundary condition. Next, we use an example to emphasize that the commonly used experimental scheme which relies on the diffusion solution with an effective thermal conductivity may sometimes fails in capturing the complete physics. Correspondingly, we adapt the scheme from Ref. 19 to develop a consistent experimental procedure, and demonstrate using a virtual experiment.

### A. Gray vs. nongray models

We first consider a gray model. Expressing the governing equation (Eq.7) in a two-dimensional Cartesian coordinate[37-38] and performing the Fourier cosine transform with respect to $x$, one obtains

$$\frac{i\omega_H}{\alpha_{AC,gray}}\tilde{G}_{gray} = -\lambda^2 \tilde{G}_{gray} + \frac{\partial^2 \tilde{G}_{gray}}{\partial y^2}. \tag{17}$$

where $\tilde{G}_{gray}(\lambda) = \int_0^\infty \bar{G}_{gray}(x)\cos(\lambda x)\,dx$. Likewise, the boundary condition imposed by the heater ($|x| \leq b$, $y = 0$), and that on the far-end ($y = \infty$) are also transformed to Fourier space,

$$\tilde{q}_{net,bdy}(\lambda, y = 0) = \frac{P_{0,gray}}{2l}\frac{\sin(\lambda b)}{\lambda b}, \tag{18a}$$

$$\tilde{G}_{gray}(y = \infty) - 2\frac{2-\epsilon_{gray}}{\epsilon_{gray}}\frac{k_{AC,gray}}{C_{gray}v_{gray}}\nabla \tilde{G}_{gray}(y = \infty)\cdot \hat{\mathbf{n}} = 0. \tag{18b}$$

Following the solution strategy in Section II-C, we first find the existing general solution to the corresponding diffusion equation.[37-38] Then we replace $k_{bulk}$ in this solution with $k_{AC,gray}$ to obtain the general solution of Eq. 17, with undetermined coefficients. At last, we apply the boundary conditions (Eqs. 18a and 18b) to solve for the undetermined coefficients and obtain the solution,

$$\tilde{G}_{gray}(\lambda, y) = \frac{1}{\tilde{\mu}_{gray}}\frac{P_{0,gray}}{2l}\frac{\sin(\lambda b)}{\lambda b}e^{-\tilde{u}_{gray}y}, \tag{19}$$

where $\tilde{\mu}_{gray} = k_{AC,gray}\tilde{u}_{AC,gray}/C_{gray}v_{gray}$ and $\tilde{u}_{AC,gray} = \sqrt{i\omega_H/\alpha_{AC,gray} + \lambda^2}$.

One further obtains $\tilde{E}_{b,gray,bdy}$ using Eq. 13, and $\Delta \tilde{T}_{BTE,gray}$ ($=\tilde{T}_{bdy} - T_{ref.}$) using Eq. 14. We then transform it back to real space using the inverse Fourier cosine transform,

$$\Delta \bar{T}_{BTE,gray}(x, y = 0) = \frac{P_{0,gray}}{\pi l} \int_0^\infty \frac{1}{k_{AC,gray}\tilde{u}_{AC,gray}} \frac{\sin(\lambda b)}{\lambda b} \cos(\lambda x)\, d\lambda$$
$$+ 2\frac{2-\epsilon_{gray}}{\epsilon_{gray}} \frac{1}{C_{gray}v_{gray}} \frac{P_{0,gray}}{2bl} \frac{1}{2}[\text{sgn}(x+b) - \text{sgn}(x-b)], \quad (20)$$

where $\Delta \bar{T}_{BTE,gray}(x) = \frac{2}{\pi}\int_0^\infty \Delta \tilde{T}_{BTE,gray}(\lambda)\cos(\lambda x)\, d\lambda$, and sgn denotes the Sign function. We note that the 1st term on RHS of Eq. 20 recovers the form of the corresponding diffusion solution with the only difference of replacing $k_{bulk}$ with $k_{AC,gray}$. The 2nd term arises from the jump boundary condition and introduces the nonlocal effect. In the limit of $\lambda b \ll 1$, the integral on the right-hand side of Eq. 20 can be approximated by

$$\frac{P_{0,gray}}{\pi l}\int_0^\infty \frac{1}{k_{AC,gray}\tilde{u}_{AC,gray}} \frac{\sin(\lambda b)}{\lambda b}\cos(\lambda x)\, d\lambda \approx \frac{P_{0,gray}}{\pi l k_{AC,gray}} K_0(u_{AC,gray}r), \quad (21)$$

which recovers the well-known solution describing a semi-infinite medium heated by an infinitely-narrow heater.[10, 32]

Next, we solve the nongray problem. One simply replaces the subscript, *gray*, with $\omega$ in Eqs. 17-19. One caveat in practical experiments, though, is that, instead of a spectral flux boundary condition, a total flux boundary condition due to Joule heating, $q_{net,bdy} = \bar{q}_{net,bdy}e^{i\omega_H t} = \int_0^\infty \bar{q}_{net,\omega,bdy}\,d\omega\, e^{i\omega_H t}$, is given, where $\bar{q}_{net,\omega,bdy} = P_{0,\omega}/(2bl)$ and $\bar{q}_{net,bdy} = P_0/2bl$. As a result, the heater temperature cannot be obtained by simply replacing the subscript, *gray*, with $\omega$ in Eq. 20.

With details deferred to Appendix E, we are able to address this problem, and obtain the nongray solution,

$$\Delta \bar{T}_{BTE,nongray}(x, y=0) = \frac{P_0}{\pi l}\int_0^\infty \frac{1}{\int_0^\infty \frac{\tilde{\mu}_\omega}{1+2\frac{2-\epsilon_\omega}{\epsilon_\omega}\tilde{\mu}_\omega}C_\omega v_\omega d\omega} \frac{\sin(\lambda b)}{\lambda b}\cos(\lambda x)\, d\lambda. \quad (22)$$

### B. Amplitude and phase lag of the surface temperature response

To facilitate the discussion in the following, we separate the amplitude and the phase lag of the surface temperature response to the periodic surface heating,

$$\frac{\Delta T_{BTE,avg.}}{q_{net,bdy}} = \frac{\Delta \bar{T}_{BTE,avg.}}{\bar{q}_{net,bdy}} = A e^{i\phi_{lag}}, \quad (23)$$

where we omit the common factor, $e^{i\omega_H t}$, for both the numerator and the denominator of the expression between the two equal signs. Here $\Delta \bar{T}_{BTE,avg.}$ denotes the average of the surface temperature response across the heater width. Equation 23 in fact define the thermal resistance per unit area (with SI units of [m$^2$-K/W]),

$$R''_{BTE} = \frac{\Delta \bar{T}_{BTE,avg.}}{\bar{q}_{net,bdy}} = \frac{2b}{\pi}\int_0^\infty \frac{1}{k_{AC,gray}\tilde{u}_{AC,gray}} \frac{\sin^2(\lambda b)}{(\lambda b)^2}\, d\lambda + 2\frac{2-\epsilon_{gray}}{\epsilon_{gray}}\frac{1}{C_{gray}v_{gray}}, \quad (24)$$

where we use the gray model (Eq. 20) to compute $\Delta \bar{T}_{BTE,avg.}$ to better correlate the frequency and the nonlocal effects to their corresponding mathematical expressions, as will be clear in the following discussion. On the other hand, to mimic the real measurements, in next section we will use the nongray model (Eq. 22) to compute $\Delta \bar{T}_{BTE,avg.}$ for the virtual experiments.

We first examine the amplitude, $A$ (or, equivalently $|R''_{BTE}|$, red solid lines in Fig. 3a). We refer the 1$^{st}$ term on RHS of Eq. 24 as $R''_{BTE,freq.}$ (black dot-dashed line in Figs. 3a), and the 2$^{nd}$ as $R''_{BTE,nonlocal}$ (black dashed line). In the low frequency limit ($\gamma_{gray} \ll 1$), it is apparent that $k_{AC,\omega}$ (Eq. 9) reduces to $k_{bulk}$, and thus $R''_{BTE,freq.}$ recovers its Fourier-law counterpart,

$$R''_{FL} = \frac{2b}{\pi} \int_0^\infty \frac{1}{k_{bulk} \tilde{u}_{bulk}} \frac{\sin^2(\lambda b)}{(\lambda b)^2} d\lambda, \tag{25}$$

where $\tilde{u}_{bulk} = \sqrt{i\omega_H/\alpha_{bulk} + \lambda^2}$, as confirmed by the black dot-dashed line and red dashed line in Fig. 3a.

On the other hand, $R''_{BTE,nonlocal}$ introduces a nonlocal effect, which had been discussed in the context of steady-state line-heating of a semi-infinite substrate.[39] We now rigorously analyze this effect for the periodic line-heating scenario ($|\tilde{u}_{bulk}b| \ll 1$), in which case Eq. 25 simplifies to

$$R''_{FL,line-heating} \approx \frac{1}{C_{gray}v_{gray}} \frac{3}{Kn_{gray}} \ln\left(\frac{2Kn_{gray}}{\sqrt{3i\gamma_{gray}}}\right), \tag{26}$$

which gradually vanishes as the decrease of $Kn_{gray}$, and thus implies an infinite heating flux for a finite temperature gradient when $Kn_{gray}$ approaches infinity, i.e., $b$ approaches zero for a fixed $\Lambda_{gray}$.[40] The BTE solution (Eq. 24) circumvents this nonphysical problem by its second term, $R''_{BTE,nonlocal} = 2\frac{2-\epsilon_{gray}}{\epsilon_{gray}} \frac{1}{C_{gray}v_{gray}}$ (black dashed line), in which we set $\epsilon_{gray} = 1$ in Fig. 3.

We note that this nonlocal effect, which originates from the jump boundary condition (Eq. 12), should be distinguished from the conventional size effect, in which the phonon MFP is truncated by real physical boundaries, e.g. the diameter of nanowires,[41] the grain size of nanocrystalline materials,[42] or the thickness of thin films.[15]

We then push to the high frequency limit ($\gamma_{gray} \gg 1$) to discuss a heating-frequency effect. Assuming planar heating ($|\tilde{u}_{bulk}b| \gg 1$), one simplifies Eq. 25 to

$$R''_{FL,planar-heating} \approx \frac{\sqrt{3}}{C_{gray}v_{gray}} \frac{1}{\sqrt{i\gamma_{gray}}}, \tag{27}$$

which, again, implies a vanished thermal resistance and thus an infinite heating flux for a finite temperature gradient when $\gamma_{gray}$ approaches to infinity. In this scenario, the BTE solution (Eq. 24) addresses this nonphysical shortcoming even without the help of its second term, $R''_{BTE,nonlocal}$ (black dashed line), because in this limit, its first term simplifies to

$$R''_{BTE,freq.,planar-heating} \approx \frac{\sqrt{3}}{C_{gray}v_{gray}}, \tag{28}$$

which is independent of the heating frequency, $\omega_H$, and thus leads to a plateau (black dot-dashed line in Fig. 3a). This independence of $\omega_H$ could be explained by the fact that the penetration depth (Eq. 16) saturates to a constant value, $Lp_{BTE} \to \frac{2}{\sqrt{3}}\Lambda$ as $\gamma_{gray} \to \infty$;[19, 24] in contrast, its Fourier-law counterpart, $Lp_{FL}$, (the equation beneath Eq. 16) keeps decreasing to zero.

Next, in Fig. 3b we plot the phase lag, $\phi_{lag}$, as a function of $\gamma_{gray}$ for various $Kn_{gray}$ using both the BTE solutions (Eq. 24; solid lines) and the corresponding Fourier-law solutions (Eq. 25; dashed lines). Regardless of the values of $Kn_{gray}$, two general features can be clearly recognized from Fig. 3b. First, the Fourier-law solutions (dashed lines) recover both the infinitely-narrow-line heating limit ($\phi_{lag} = 0°$) as $\gamma_{gray} \to 0$) and the planar heating limit ($\phi_{lag} = -45°$) as $\gamma_{gray} \to \infty$), as they should.[43] Second, while the BTE solutions recover their corresponding Fourier-law solutions in the low $\gamma_{gray}$ regime, as $\gamma_{gray}$ keeps increasing, they roll back to $0°$, which is a key distinction as compared to the Fourier-law solutions. This rolling back is a direct consequence of $k_{AC,gray} = k_{bulk}/(1 + i\omega_H \tau_{gray})$, in which the pre-factor, $(1 + i\omega_H \tau_{gray})$, causes this additional phase variation. In addition to these common features, Fig. 3b also shows that as $Kn_{gray}$ increases, some of the BTE solutions (blue and green) cannot recover the diffusive planar-heating limit ($\phi_{lag} = -45°$). This is because as the width of the heater keeps decreasing, it requires larger $\omega_H$ to further reduce the penetration depth to reach the diffusive planar-heating limit. However, the rolling-back effect of $k_{AC}$ starts to dominate before that limit is reached.

## C. Virtual experiments to extract both scattering and transmission parameters

In practical measurements, it is usually applied without justification to fit non-diffusive experimental data to the diffusion equation, with an effective thermal conductivity, $k_{eff}$. In the following, we show that this $k_{eff}$- approach may lead to non-negligible errors in some scenarios,[25-26, 44] which motivates the necessity of implementing the ballistic $3\omega$ method.

Figure 4 illustrates this problem by comparing various fittings (lines) using this $k_{eff}$- approach to the "true physics" (symbols) generated using Eq. 20 with $Kn_{gray} = 1$ and $\gamma_{gray} = 10^{-4}$ for intrinsic silicon at 300 K. While in a standard routine of the classic $3\omega$ experiment, one measures only the temperature of the heater, here we examine the full surface temperature distribution. The comparisons show that while one trial with $k_{eff.} = k_{bulk}$ (blue) fails in capturing the heater temperature, another trial with $k_{eff.} = 0.6 \times k_{bulk}$ (red) cannot capture the surface temperature profile other than the heater region. If one performs a brute-force approach to fit the "true physics" with a "best-fit" $k_{eff.} = 0.7 \times k_{bulk}$ (green), it succeeds in neither region. This failure originates from the 2nd term on RHS of Eq. 20, which is estimated to be up to 40% of the 1st term in this scenario. In this case, it is impossible to cast the 2nd term and thus the BTE solution (Eq. 20) into the form of its corresponding Fourier-law solution using a $k_{eff.}$.

The example above highlights the inconsistency problem of this $k_{eff.}$- approach. To address this problem, we propose a consistent approach that fits "experimental data" with the ballistic $3\omega$ model. We adapt the literature scheme to extract from the $\phi_{lag} - \omega_H$ relation not only the phonon scattering information,[19] but also the transmission coefficient of phonons from the heater to the sample. In this strategy, we assume the phonon dispersion (and thus the group velocity and heat capacity) is obtained either experimentally or theoretically in advance. In addition, we choose a phenomenological scattering model, $\tau_\omega^{-1} = D\omega^n$, where $D$ and $n$ are free parameters indicating the dominant scattering mechanism. For example, Umklapp scattering is dominant if $D \propto T$ and $n \approx 2$.

We generate synthetic "experimental" data of the $\phi_{lag} - \omega_H$ relation using Eqs. 22 and 23, including $\pm 2\%$ of "experimental noise" (open circles in Fig. 5a), in which we combine a lumped single-branch BvK dispersion[19, 30] with various scattering mechanisms, including a strong Umklapp scattering, $\tau_{umkl.}^{-1} = 1.53 \times 10^{-19}\omega^2 T$ with $T = 300$ K, a weak impurity scattering, $\tau_{imp.}^{-1} = 2.54 \times 10^{-45}\omega^4$, and a boundary scattering, $\tau_{bdy.}^{-1} = v_\omega/(5.7 \times 10^{-3})$, all with SI units of [s$^{-1}$]. To simplify the fitting process, here we assume a gray transmission coefficient between the gold heater and the silicon substrate, $\epsilon_\omega = \epsilon_{htr-sub.} = 0.90$ according to the DMM model.[29, 45]

We fit Eqs. 22 and 23 to the synthetic "experimental" data in Fig. 5a to obtain both scattering and transmission parameters. In the low $\omega_H$ regime before the minimum phase lag, $\phi_{lag,min} = -45°$, corresponding to a cutoff heating-frequency of $\omega_{H,c} \approx 10^6$ rad/s, the nonlocal effect is negligible. Therefore, we omit the term involving $\epsilon_\omega$ in Eq. 22, and fit for the two scattering parameters (red line in Fig. 5a), which yields $D = 1.74 \times 10^{-19} T$ and $n = 2.00$, in good agreement with the actual Umklapp parameters used above to generate the "experimental" data. This result also confirms that the Umklapp process is the dominant scattering mechanism for silicon at 300 K. With $D$ and $n$ obtained, we reconstruct the MFT distribution function (open red

circles in Fig. 5b),[19]

$$K_\tau(\tau_\omega) = -\frac{1}{3}C_\omega v_\omega^2 \tau_\omega \left(\frac{d\tau}{d\omega}\right)^{-1}, \qquad (30)$$

which agrees within 8% with the "true" distribution function (solid black line in Fig. 5b) computed using the "true" scattering parameters to generate the "experiments" (open circles in Fig. 5a). In the high $\omega_H$ regime ($>\omega_{H,c}$), we fit for $\epsilon_{htr-sub.}$ (blue line in Fig. 5a). In this regime, the term involving $\epsilon_\omega$ in Eq. 22, which is simplified to a gray $\epsilon_{htr-sub.}$, is non-negligible, if not dominant. This one-parameter fitting yield $\epsilon_{htr-sub.} = 0.86$, which agrees to better than 5% with the "true value" simulated using the DMM model above. In Fig. 5c, we replace the gold heater with various other metals to further validate this fitting scheme.

## VI. Conclusions and discussion

With the phonon-photon analogy, we exploited the P$_1$-approximation from thermal radiation to transform the phonon BTE under RTA to a diffusion equation (Eq. 7) with an AC thermal conductivity (Eq. 9). We further applied the jump boundary condition to take into account the nonlocal effect. With this framework, we can map any existing Fourier-law solution under periodic heating to the corresponding BTE solution. We verified this framework using existing analytical solutions and quantified its applicable region using numerical solutions. As an example to implement this framework, we generalize the classic $3\omega$ method from diffusive to quasi-ballistic. To address the inconsistency problem of prevailing experiments which fit quasi-ballistic data with Fourier-law models, we proposed experimental scheme to extract both scattering parameters and the transmission coefficient in a consistent way and demonstrated using virtual experiments.

Existing electronic technologies, e.g. the radio frequency (RF) technology, employs devices that can generate and detect electrical signals beyond GHz level. Therefore, conducting $3\omega$ measurements above GHz is not a problem, although issues such as the power-carrying capability of these devices and the inductive and/or capacitive impedance matching need to be carefully considered. On the other hand, we do recognize several aspects for further development of this framework. First, although we use the $3\omega$ method to illustrate how to use this framework to generalize existing experimental techniques from diffusive to quasi-ballistic, it can be applied to other techniques relying on periodic heating, such as the TDTR and FDTR. Second, in the synthetic "experiments", we pre-assume the form of the phonon scattering relation and a gray phonon transmission coefficient for the sake of simplicity. Advanced numerical scheme is desired to extract the $\tau_\omega - \omega$ relation and the spectral transmission coefficient from the measure $\phi_{lag} - \omega_H$ relation by solving a Fredholm integral equation.[46] At last, extending this BTE scheme to multilayer structures, such as microchips, by a reliable numerical formalism such as the scattering matrix method[38] is desired.

**Acknowledgements**

This work was supported in part by National Natural Science Foundation of China (52376051). We thank Chris Dames and Gang Chen for helpful discussions.

**Tables and Figures**

| Geometry | Fourier solutions $(\bar{T}^* = \frac{\bar{T}-T_2}{\Delta T})$ | BTE solutions $(\bar{G}^* = \frac{\bar{G}-4E_{b,2}}{4\Delta E})$ |
|---|---|---|
| Infinite parallel plates 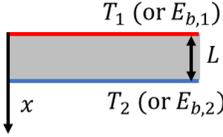 $T_1$ (or $E_{b,1}$), $L$, $T_2$ (or $E_{b,2}$), $x$ | $\dfrac{e^{u(L-x)} - e^{-u(L-x)}}{e^{uL} - e^{-uL}}$ | $\dfrac{C_{p,1}e^{u_{AC}(L-x)} - C_{p,2}e^{-u_{AC}(L-x)}}{C_{p,1}^2 e^{u_{AC}L} - C_{p,2}^2 e^{-u_{AC}L}}$ |
| Infinitely long concentric cylinders 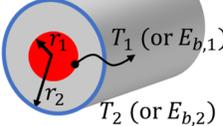 $T_1$ (or $E_{b,1}$), $T_2$ (or $E_{b,2}$), $r_1$, $r_2$ | $\dfrac{I_0(ur_1)K_0(ur) - K_0(ur_2)I_0(ur)}{K_0(ur_1)I_0(ur_2) - K_0(ur_2)I_0(ur_1)}$ | $\dfrac{C_{c,2}K_0(u_{AC}r) - C_{c,4}I_0(u_{AC}r)}{C_{c,2}C_{c,3} - C_{c,1}C_{c,4}}$ |
| Concentric spheres 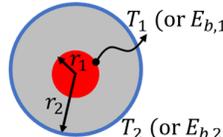 $T_1$ (or $E_{b,1}$), $T_2$ (or $E_{b,2}$), $r_1$, $r_2$ | $\dfrac{r_1}{r}\dfrac{e^{u(r_2-r)} - e^{-u(r_2-r)}}{e^{u(r_2-r_1)} - e^{-u(r_2-r_1)}}$ | $\dfrac{r_1}{r}\dfrac{C_{s,4}e^{u_{AC}(r_2-r)} - C_{s,3}e^{-u_{AC}(r_2-r)}}{C_{s,2}C_{s,4}e^{u_{AC}(r_2-r_1)} - C_{s,1}C_{s,3}e^{-u_{AC}(r_2-r_1)}}$ |

Table I. Representative examples to compare the BTE solutions (right column; this work) to their Fourier-law counterparts (left column). The former shares the same mathematical structure with the latter, except that $k_{bulk}$ is replaced by $k_{AC}$ (Eq. 9), and thus $u_{bulk}$ is replaced by $u_{AC} = \sqrt{i\omega_H/\alpha_{AC}}$. For clarity, we omit the subscript, $\omega$, of $G_\omega$, $E_{b,\omega}$ and $u_{AC,\omega}$; and *bulk* of $u_{bulk}$. Coefficients of the BTE solutions, $C_{geom.,n}$ (Table S1 of Appendix C), are quite different as compared to those of the Fourier-law solutions, due to the jump boundary condition (Eq. 13). Here $T_1 = \Delta T e^{i\omega_H t} + T_2$ and $E_{b,1} = \Delta E e^{i\omega_H t} + E_{b,2}$. $I_0$ and $K_0$ are the modified Bessel function of the first and second kind, respectively.

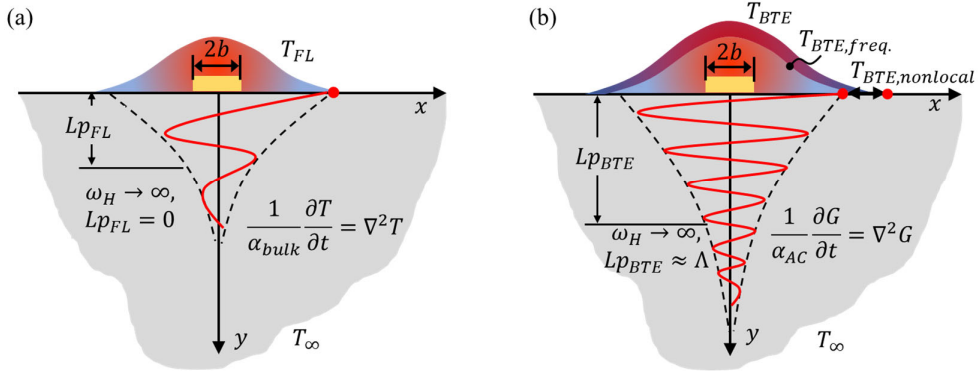

Fig. 1. Schematic of an example to implement the framework: generalizing the $3\omega$ method from (a) diffusive to (b) quasi-ballistic (see Section V for details). Here $T_{BTE} = T_{BTE,freq.} + T_{BTE,nonlocal}$, where $T_{BTE,freq.}$ shares the same mathematical form with $T_{FL}$, except that $k_{bulk}$ is replaced by $k_{AC}$ (Eqs. 24 and 25). This is because the same equations ($T$ vs. $G$) have the same solutions. Here $G \propto T_{BTE,freq.}$ in a linearized approximation. The nonlocal effect, $\Delta T_{nonlocal}$, is approximated using the jump boundary condition (Eq. 24). Note that while the penetration depth of the Fourier-law solution, $Lp_{FL} = \sqrt{2\alpha_{bulk}/\omega_H}$, gradually diminishes as the increase of the heating frequency, $\omega_H$, that of the BTE solution ($Lp_{BTE}$; Eq. 16) decreases to a constant that is proportional to the phonon mean free path, $\Lambda$.[19, 24]

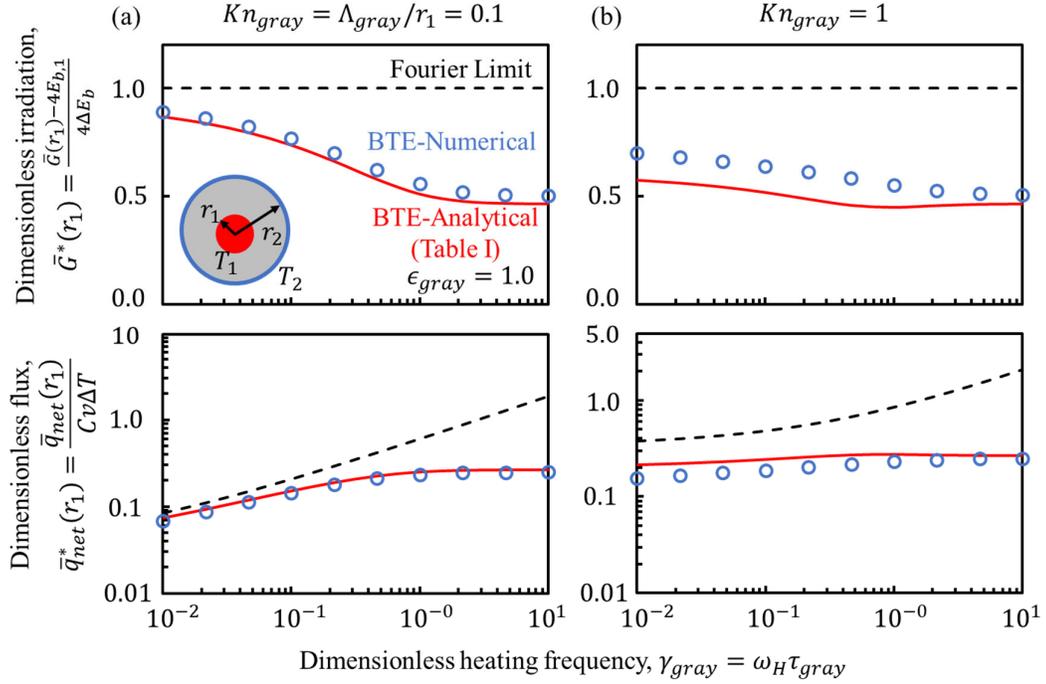

Fig. 2. Numerical verification (symbols) of the analytical BTE solution (lines) of the dimensionless phonon irradiation (upper row) and heat flux (lower row) of the concentric spherical problem in Table I with $Kn_{gray} = \Lambda_{gray}/r_1 = 0.1$ (a) and 1 (b), respectively. The Fourier limits (dashed) are also shown for comparison. Analytical solutions agree well with the numerical results for small $Kn_{gray}$ ($\leq 0.1$) regardless of the heating frequency, $\gamma_{gray}$. As $Kn_{gray}$ increases to unity, analytical solutions start to deviate from the numerical solutions, in particular in the small $\gamma_{gray}$ regime. The discrepancy will increase as $Kn_{gray}$ increases further to be beyond unity, which, however, is less practical in real measurements.

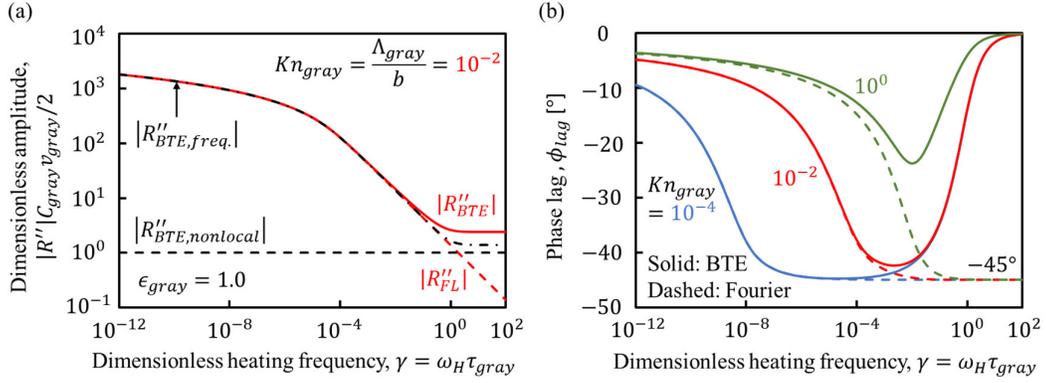

Fig. 3. Amplitude (a) and phase lag (b) of the surface temperature response: BTE solutions (Eq. 24; solid lines) vs. Fourier limits (blue, red, and green dashed lines). In the small $\gamma_{gray}$ limit, both the amplitude and the phase lag of the BTE solution reduce to their corresponding Fourier limits. Note here $\phi_{lag} = 0°$ and $\phi_{lag} = -45°$ correspond to the diffusive infinitely-narrow line- and planar-heating limits. As $\gamma_{gray}$ keeps increasing, however, the BTE solution gradually deviates from the Fourier limit. The rolling-back of $\phi_{lag}$ to $0°$ in the large $\gamma_{gray}$ regime is mainly due to $k_{AC}$ (Eq. 9).

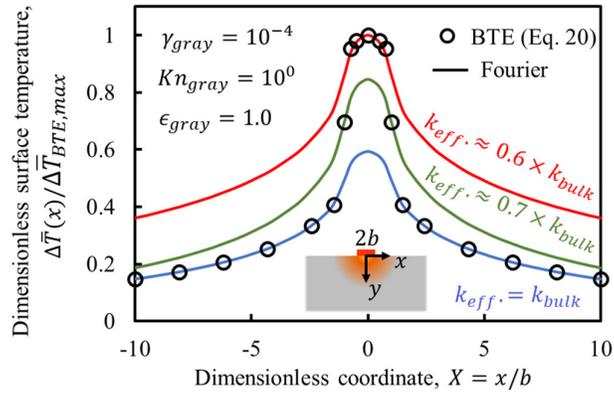

Fig. 4. An example to show the failure of the $k_{eff.}$- approach. Fitting the BTE data (symbols; Eq. 20) to the Fourier-law solution (lines) cannot capture the full surface temperature distribution, regardless of the free parameter, $k_{eff.}$. This failure motivates the necessity of an experimental scheme to fit the non-diffusive experimental data with a BTE model (Fig. 5).

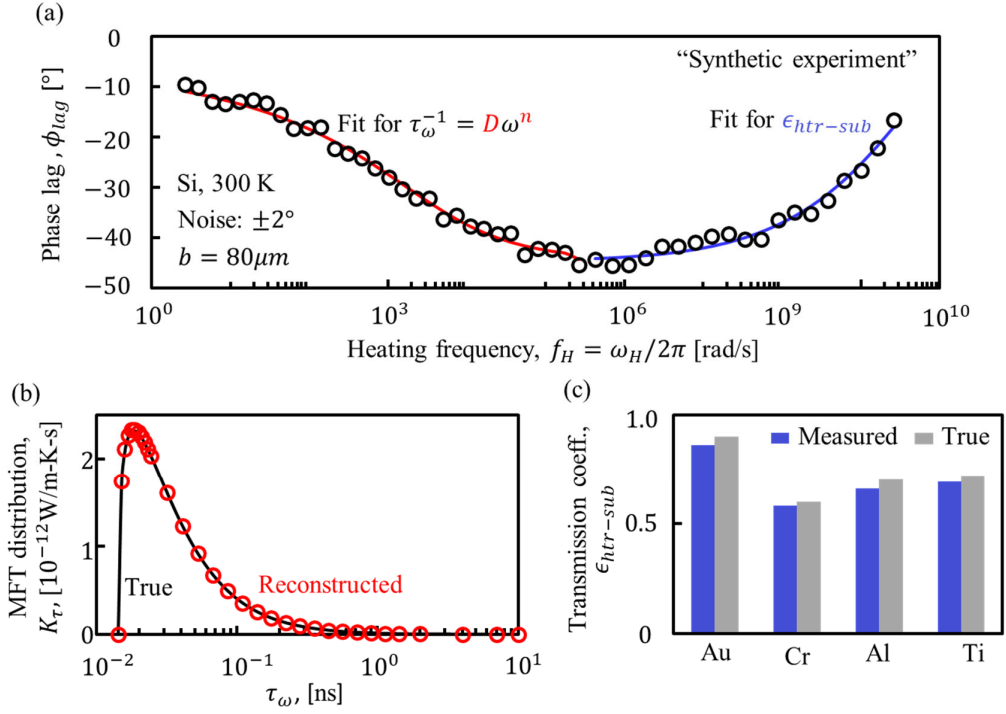

Fig. 5. Proposed scheme to measure both scattering and transmission parameters by fitting non-diffusive data to a BTE model. (a) Synthesized data (Eqs. 22 and 23), with $\pm 2\%$ of "experimental noise" (symbols). Scattering parameters, $D$ and $n$, and the transmission coefficient, $\epsilon_{htr-sub}$, are determined by fitting Eq. 23 to the low- and high-frequency data, respectively (red and the blue lines). (b) Reconstructed mean free time (MFT) distribution (symbols) using "measured" scattering parameters agrees within 8% with the true distribution (line). (c) "Measured" transmission coefficients agree within 5% with true values of various metal heaters.

APPENDICES

APPENDIX A: Phonon distribution function under periodic heating

APPENDIX B: Detailed derivation of the decoupled $G_\omega$ and $\boldsymbol{q}_{net,\omega}$ equations

APPENDIX C: Coefficients of the solutions of the representative examples in Table I

APPENDIX D. Comparison to results of Ref. 19.

APPENDIX E: Detailed derivation of the nongray ballistic $3\omega$ model

APPENDIX A: Phonon distribution function under periodic heating

Following Refs. 2 and 5, we derive the AC thermal conductivity. Recall BTE under RTA,

$$\frac{\partial f_\omega}{\partial t} + v_\omega \hat{\mathbf{s}} \cdot \nabla f_\omega = -\frac{f_\omega - f_\omega^0}{\tau_\omega}. \tag{A1}$$

If the deviation function, $g_\omega = f_\omega - f_\omega^0$, and its gradient in real space, $\nabla g_\omega$, are much smaller than $f_\omega^0$ and $\nabla f_\omega$, respectively, one may replace $\nabla f_\omega$ in Eq. A1 with $\nabla f_\omega^0$, and thus Eq. A1 is transformed to

$$\frac{\partial f_\omega}{\partial t} + v_\omega \hat{\mathbf{s}} \cdot \nabla f_\omega^0 = -\frac{f_\omega - f_\omega^0}{\tau_\omega}. \tag{A2}$$

According to the linear response theorem, in response to the AC stimulus, $T = \bar{T}(x)e^{i\omega_H t}$, we have $\partial f_\omega/\partial t = i\omega_H f_\omega$. Substituting it to Eq. A2, one obtains

$$f_\omega = \frac{1}{1+i\omega_H \tau_\omega}(f_\omega^0 - v_\omega \tau_\omega \hat{\mathbf{s}} \cdot \nabla f_\omega^0), \tag{A3}$$

in which a suppression factor, $1/(1 + i\omega_H \tau_\omega)$, is introduced. As argued in the following, this suppression factor will go all the way through the derivation of the heat flux and appear in the AC thermal conductivity.

Substituting Eq. A3 into the definition of the heat flux, which is assumed to be along the *x*-axis for simplicity,

$$q_{x,\omega} = \sum_{pol.} \left[\frac{1}{V}\sum_{\mathbf{k}} v_\omega \hat{\mathbf{s}} \cdot \hat{\mathbf{x}} \hbar \omega f_\omega\right], \tag{A4}$$

one can drop the $f_\omega^0$ term in Eq. A3 according to the symmetry argument,[6] and obtain the expression of the spectral AC thermal conductivity,

$$k_{AC,\omega} = \frac{C_\omega v_\omega^2 \tau_\omega/3}{1+i\omega_H \tau_\omega}, \tag{A5}$$

where the spectral heat capacity, $C_\omega = \hbar\omega D(\omega)\frac{\partial f_\omega^0}{\partial T}$, and $D(\omega)$ is the phonon density of states.

## B. Detailed derivation of the decoupled $G_\omega$ and $\boldsymbol{q}_{net,\omega}$ equations

We show how to decouple $G_\omega$ and $\boldsymbol{q}_{net,\omega}$ from Eqs. 3 and 5 of the main text. The math is very similar to that used to decouple the electrical and magnetic fields from the Maxwell equations to obtain the wave equation.[27]

First, we derive the $G_\omega$ equations (Eqs. 6 and 7 of the main text). Taking the divergent of Eq. 3 of the main text, one obtains

$$\nabla \cdot \frac{\partial \boldsymbol{q}_{net,\omega}}{\partial t} + \frac{v_\omega}{3} \cdot \nabla^2 G_\omega = -\frac{\nabla \cdot \boldsymbol{q}_{net,\omega}}{\tau_\omega}. \tag{B1}$$

Taking the time derivative of Eq. 5, one arrives at

$$\frac{\partial^2 G_\omega}{\partial t^2} + \nabla \cdot (v_\omega \frac{\partial \boldsymbol{q}_{net,\omega}}{\partial t}) = 0. \tag{B2}$$

Combining Eqs. B1 and B2, and substituting $\nabla \cdot \boldsymbol{q}_{net,\omega}$ with $\partial G_\omega / \partial t$ from Eq. 5, one eliminates $\boldsymbol{q}_{net,\omega}$ and obtains Eq. 6 of the main text:

$$\frac{\partial^2 G_\omega}{\partial t^2} + \frac{1}{\tau_\omega} \frac{\partial G_\omega}{\partial t} = \frac{v_\omega^2}{3} \nabla^2 G_\omega. \tag{B3}$$

According to the Linear Response Theorem, one assumes an AC response, $G_\omega(\mathbf{r}, t) = \bar{G}_\omega(\mathbf{r}) e^{i\omega_H t}$, to the periodic heating excitation with an angular frequency of $\omega_H$. Substituting this AC response to Eq. B3, one obtains

$$i\omega_H (1 + i\omega_H) \bar{G}_\omega = \frac{v_\omega^2 \tau_\omega}{3} \nabla^2 G_\omega. \tag{B4}$$

Recognizing $i\omega_H \bar{G}_\omega = \partial G_\omega / \partial t$, one obtains Eq. 7 of the main text:

$$\frac{1}{\alpha_{AC,\omega}} \frac{\partial G_\omega}{\partial t} = \nabla^2 G_\omega. \tag{B5}$$

With a similar procedure, we next show how to obtain the $\boldsymbol{q}_{net,\omega}$ equations that are not given in the main text but may be useful in other scenarios. Taking the time derivative of Eq. 3, one arrives at

$$\frac{\partial^2 \boldsymbol{q}_{net,\omega}}{\partial t^2} + \frac{v_\omega}{3} \cdot \nabla \frac{\partial G_\omega}{\partial t} = -\frac{1}{\tau_\omega} \frac{\partial \boldsymbol{q}_{net,\omega}}{\partial t}. \tag{B6}$$

Taking the divergent of Eq. 5, one obtains

$$\nabla \frac{\partial G_\omega}{\partial t} + v_\omega \cdot \nabla(\nabla \cdot \boldsymbol{q}_{net,\omega}) = 0, \tag{B7}$$

Combining Eqs. B6 and B7, one eliminates $G_\omega$ and obtain the governing equation of $\boldsymbol{q}_{net,\omega}$:

$$\frac{\partial^2 \boldsymbol{q}_{net,\omega}}{\partial t^2} + \frac{1}{\tau_\omega} \frac{\partial \boldsymbol{q}_{net,\omega}}{\partial t} = \frac{v_\omega^2}{3} \cdot \nabla(\nabla \cdot \boldsymbol{q}_{net,\omega}). \tag{B8}$$

Likewise, substituting $\boldsymbol{q}_{net,\omega}(\mathbf{r}, t) = \bar{\boldsymbol{q}}_{net,\omega}(\mathbf{r}) e^{i\omega_H t}$ to Eq. B8 and re-arrange, one obtains

$$\frac{1}{\alpha_{AC,\omega}} \frac{\partial \boldsymbol{q}_{net,\omega}}{\partial t} = \nabla(\nabla \cdot \boldsymbol{q}_{net,\omega}). \tag{B9}$$

C. Coefficients of the BTE solutions in Table I of the main text

| Geometry | General expression | $C_{geom.,1}$ | $C_{geom.,2}$ | $C_{geom.,3}$ | $C_{geom.,4}$ |
|---|---|---|---|---|---|
| Infinite parallel plates | $1 \pm 2\dfrac{2-\epsilon}{\epsilon}\mu$ | + | − | | |
| Infinitely long concentric cylinders | $I_0(u_{AC}r_j) \pm 2\dfrac{2-\epsilon}{\epsilon}\mu I_1(u_{AC}r_j)$ | $j=1$ − | $j=2$ + | | |
| | $K_0(u_{AC}r_j) \pm 2\dfrac{2-\epsilon}{\epsilon}\mu K_1(u_{AC}r_j)$ | | | $j=1$ + | $j=2$ − |
| Concentric spheres | $1 + 2\dfrac{2-\epsilon}{\epsilon}\mu(1/u_{AC}r_j \pm 1)$ | $j=1$ − | $j=1$ + | | |
| | $1 + 2\dfrac{2-\epsilon}{\epsilon}\mu(-1/u_{AC}r_j \pm 1)$ | | | $j=2$ − | $j=2$ + |

Table S1. Compact summary to find the coefficients, $C_{geom.,n}$, of the BTE solutions. For example, to find $C_{c,1}$, the 1$^{st}$ coefficient of the solution to the problem of the infinitely long concentric cylinders, one simply replaces the subscript of $r_j$ with "1", and chooses the "-" sign from the corresponding general expression in the 2$^{nd}$ column. Therefore, one determines $C_{c,1} = I_0(u_{AC}r_1) - 2\dfrac{2-\epsilon_\omega}{\epsilon_\omega}\mu I_1(u_{AC}r_1)$. Here $\epsilon$ represents the transmission coefficient from the heater to the sample in the context of phonon transport.[6, 24] and $\mu = k_{AC}u_{AC}/Cv$ For clarity, we omit the subscript, $\omega$, in all these variables.

D. Comparison to results of Ref. 19.

We show both the seemingly complicated BTE temperature and the suppression function in Ref. 19 can be significantly simplified using the AC thermal conductivity notation in this work.

In Ref. 19, the BTE temperature and the net flux of a semi-infinite medium read

$$\Delta T_{BTE,equil.} = \Delta T \frac{\int_0^\infty \frac{C_\omega v_\omega}{2}(1+de^{i\phi})e^{\frac{ia+b}{\beta v_\omega \tau_\omega}x}d\omega}{\int_0^\infty C_\omega v_\omega d\omega} \times e^{i\omega_H t}, \tag{D1}$$

$$q_{net} = \Delta T \int_0^\infty \frac{C_\omega v_\omega}{2}(1-de^{i\phi})e^{\frac{ia+b}{\beta v_\omega \tau_\omega}x}d\omega \times e^{i\omega_H t}, \tag{D2}$$

respectively, where $a = \sqrt{\frac{\gamma_\omega}{2}}\sqrt{\gamma_\omega + \sqrt{\gamma_\omega^2+1}}$, $b = \sqrt{\frac{\gamma_\omega}{2}}\sqrt{-\gamma_\omega + \sqrt{\gamma_\omega^2+1}}$, $d = -\frac{b}{(\beta\eta\gamma_\omega+a)\sin\phi - b\cos\phi}$, $\tan\phi = \frac{2\beta\eta\gamma_\omega b}{(\beta\eta\gamma_\omega)^2-a^2-b^2}$, $\beta = \frac{1}{\sqrt{3}}$, $\eta = 2$, and $\gamma_\omega = \omega_H \tau_\omega$.

Note here we believe the coefficient $\beta = \frac{2}{3}$ in Ref. 20 should be updated to $\beta = \frac{1}{\sqrt{3}}$. We further note that there is a typo in Ref. 19: the coefficient 2 in the denominator of Eq. D7 of Ref. 20 should be replaced by 1.

First, we recover the BTE temperature in Ref. 19. Starting from the 1st row of Table I, we further simplify the BTE temperature solution of the infinite parallel plates to its semi-infinite limit,

$$\Delta T_{BTE,equil.} = \frac{\int_0^\infty \bar{G}_\omega(x)d\omega}{\int_0^\infty C_\omega v_\omega d\omega} = \frac{\Delta T \int_0^\infty \frac{C_\omega v_\omega}{1+2\frac{2-\epsilon_\omega}{\epsilon_\omega}\frac{k_{AC,\omega}u_{AC,\omega}}{C_\omega v_\omega}}e^{-u_{AC,\omega}x}d\omega}{\int_0^\infty C_\omega v_\omega d\omega} \times e^{i\omega_H t}. \tag{D3}$$

Introducing the AC thermal conductivity $k_{AC,\omega} = \frac{C_\omega v_\omega^2 \tau_\omega/3}{1+i\omega_H \tau_\omega}$ (Eq. 9 of the main text), and setting $\epsilon_\omega = 1$, we find

$$\frac{C_\omega v_\omega}{1+2\frac{2-\epsilon_\omega}{\epsilon_\omega}\frac{k_{AC,\omega}u_{AC,\omega}}{C_\omega v_\omega}} = \frac{C_\omega v_\omega}{1+2\sqrt{\frac{3(i\gamma_\omega+1)}{i\gamma_\omega}}} = \frac{C_\omega v_\omega \sqrt{(i\gamma_\omega+1)i\gamma_\omega}}{\sqrt{(i\gamma_\omega+1)i\gamma_\omega}+i\gamma_\omega\beta\eta} = \frac{C_\omega v_\omega(1+de^{i\phi})}{2}, \tag{D4-a}$$

and

$$u_{AC,\omega} = \frac{\sqrt{3(i\gamma_\omega+1)i\gamma_\omega}}{v_\omega \tau_\omega} = \frac{ia+b}{\beta v_\omega \tau_\omega}, \tag{D4-b}$$

where we use $1+de^{i\phi} = \frac{2\sqrt{(i\gamma_\omega+1)i\gamma_\omega}}{\sqrt{(i\gamma_\omega+1)i\omega_H\tau+i\gamma_\omega\beta\eta}}$ and $ia+b = \sqrt{i\gamma_\omega(1+i\gamma_\omega)}$.

Then, with a very similar procedure, we recover the suppression function in Ref. 19. To proceed, we make two notes. First, for simplicity, here we consider the corresponding gray model. Second, although Ref. 19 considers only the amplitude of the suppression function, to avoid missing any information, here we consider both the real and the imaginary parts.

Following the definition of the apparent thermal conductivity, $\left.\frac{q_{net}}{\partial T/\partial x}\right|_{x=0}$, in Ref. 19, one obtains

$$\frac{k_{app,gray}}{k_{bulk}} = \frac{1-de^{i\phi}}{1+de^{i\phi}} \frac{3\beta}{2(ia+b)}, \tag{D5}$$

Again, introducing the AC thermal conductivity (Eq. 9 of the main text), one can indeed clean up these parameters, $a$, $b$, $d$, and $\phi$, and arrives at

$$\frac{k_{app,gray}}{k_{bulk}} = \frac{2i\gamma_{gray}\beta}{(ia+b)} \frac{3\beta}{2(ia+b)} = \frac{3i\gamma_{gray}\beta^2}{(ia+b)^2} = \frac{1}{1+\omega_H \tau_{gray}}, \tag{D6}$$

where we use $\frac{1-de^{i\phi}}{1+de^{i\phi}} = \frac{2i\gamma_{gray}\beta}{(ia+b)}$ and $ia + b = \sqrt{i\gamma_{gray}(1 + i\gamma_{gray})}$.

E. Detailed derivation of the nongray ballistic $3\omega$ model

We re-write Eqs. 17 - 19 to their nongray form (with subscript, *gray*, replaced with $\omega$). With $\tilde{G}_\omega$ solved, one obtains

$$\tilde{E}_{b,\omega,bdy} = \frac{1}{4}\frac{P_{0,\omega}}{2l}\frac{\sin(\lambda b)}{\lambda b}\left[\frac{1}{\tilde{\mu}_\omega} + 2\frac{2-\epsilon_\omega}{\epsilon_\omega}\right] \tag{E1}$$

using Eq. 13. Linearizing it using Eq. 14, one arrives at

$$\frac{\tilde{\mu}_\omega}{1+2\frac{2-\epsilon_\omega}{\epsilon_\omega}\tilde{\mu}_\omega}C_\omega v_\omega \Delta\tilde{T}_{BTE,nongray}(\lambda) = \frac{P_{0,\omega}}{2l}\frac{\sin(\lambda b)}{\lambda b}. \tag{E2}$$

where $\Delta\tilde{T}_{BTE,nongray}(\lambda) = \tilde{T}_{bdy} - T_{ref.}$. Integrating both sides of Eq. E2, one obtains

$$\Delta\tilde{T}_{BTE,nongray}(\lambda) = \frac{P_0}{2l}\frac{\sin(\lambda b)}{\lambda b}\frac{1}{\int_0^\infty \frac{\tilde{\mu}_\omega}{1+2\frac{2-\epsilon_\omega}{\epsilon_\omega}\tilde{\mu}_\omega}C_\omega v_\omega d\omega}. \tag{E3}$$

Note here we exploit the total flux boundary condition, $q_{net,bdy} = \int_0^\infty \bar{q}_{net,\omega,bdy}d\omega\, e^{i\omega_H t}$, where $\bar{q}_{net,\omega,bdy} = \frac{P_{0,\omega}}{2bl}$. Performing the inverse Fourier cosine transform, one obtains Eq. 22 of the main text.